\documentclass[preprint2]{aastex}
\usepackage{ulem}

\def\Ms{$M_{\odot}$}

\shorttitle{The Three New Pulsars Discovered by FAST in the Globular Cluster NGC 6517}
\shortauthors{Pan et al.}

\begin{document}

\title{Three Pulsars Discovered by FAST in the Globular Cluster NGC 6517 with a Pulsar Candidate Sifting Code Based on Dispersion Measure to Signal-to-Noise Ratio Plots}

\author{Zhichen Pan\altaffilmark{1,2,3,*}, Xiaoyun Ma\altaffilmark{1}, Lei Qian\altaffilmark{1,2,3}, Lin Wang\altaffilmark{1,2,3}, Zhen Yan\altaffilmark{4}, Jintao Luo\altaffilmark{5}, Scott M. Ransom\altaffilmark{6}, Duncan R. Lorimer\altaffilmark{7,8}, Peng Jiang\altaffilmark{1,2,3}}

\affil{$^1$National Astronomical Observatories, Chinese Academy of Sciences, 20A Datun Road, Chaoyang District, Beijing, 100101, China}
\email{panzc@bao.ac.cn}
\affil{$^2$CAS Key Laboratory of FAST, National Astronomical Observatories, Chinese Academy of Sciences, Beijing 100101, China}
\affil{$^3$College of Astronomy and Space Sciences, University of Chinese Academy of Sciences, Beijing 100049, China}
\affil{$^4$Shanghai Astronomical Observatories, Chinese Academy of Sciences, Shanghai, 20030}
\affil{$^5$National Time Service Center, Chinese Academy of Sciences, Xi'an, 710600}
\affil{$^6$National Radio Astronomy Observatory, Charlottesville, VA 22903, USA}
\affil{$^7$Department of Physics and Astronomy, West Virginia University, Morgantown, WV 26506, USA}
\affil{$^8$Center for Gravitational Waves and Cosmology, West Virginia University, Chestnut Ridge Research Building, Morgantown, WV 26505}

\begin{abstract}
  We report the discovery of three new pulsars in the Globular Cluster (GC) NGC6517, namely NGC 6517 E, F, and G, made with the Five-hundred-meter Aperture Spherical radio Telescope (FAST).
  The spin periods of NGC 6517 E, F, and G are 7.60~ms, 24.89~ms, and 51.59~ms, respectively.
  Their dispersion measures are 183.29, 183.713, and 185.3~pc~cm$^{-3}$, respectively, all slightly larger than those of the previously known pulsars in this cluster.
  The spin period derivatives are at the level of 1$\times$10$^{-18}$~s~s$^{-1}$, which suggests these are recycled pulsars.
  In addition to the discovery of these three new pulsars, we updated the timing solutions of the known isolated pulsars, NGC 6517 A, C, and D.
  The solutions are consistent with those from Lynch et al. (2011) and with smaller timing residuals.
  From the timing solution, NGC 6517 A, B (position from Lynch et al. 2011), C, E, and F are very close to each other on the sky and only a few arcseconds from the optical core of NGC 6517.
  With currently published and unpublished discoveries, NGC6517 now has 9 pulsars, ranking 5$^{th}$ of the GCs with the most pulsars.
  The discoveries take advantage of the high sensitivity of FAST
  and a new algorithm used to check and filter possible candidate signals.
\end{abstract}


\section{Introduction}

Till the submission of this paper, 221 pulsars have been discovered in 36 globular clusters\footnote{http://www.naic.edu/$\sim$pfreire/GCpsr.html,
the average number of pulsars in GCs now is larger than 1 (157 GCs in total, Harris 1996, 2010; or 160 GCs, https://people.smp.uq.edu.au/HolgerBaumgardt/globular/orbits.html),
} (GCs).
A hundred and seventeen of these pulsars, or about 53\% of the total, are gethered in the GCs those have 9 or more pulsars.
There are 39 pulsars in Terzan 5, 27 in 47 Tucanae (hereafter 47 Tuc), 14 in M28, 10 in NGC6624, 9 in NGC6752, 9 in NGC6517 (including the new pulsars in this paper), and 9 in M15.
Previous studies on the predictions of the number of pulsars in GCs also show the similar bias that the numbers of GC pulsars can be significantly different from each GC
and a much larger number of pulsars should be discovered.
These studies usually focused on GCs with pulsars more than 5 pulsars.
Bagchi et al. (2011) used a Monte Carlo simulation to model the luminosity distribution of recycled pulsars in GCs.
They gave the predictions for 10 GCs under different simulation parameters,
as Terzan 5, 47 Tuc, M28, M15, and NGC 6440 are the five with the largest number of pulsars.
Chennamangalam et al. (2013) used Bayesian methods to explore the mean and standard deviation of the luminosity function and total number of pulsars in GCs.
While they only applied the simulation to Terzan 5, 47 Tuc and M28,
the result shows that M28 may have larger number of pulsars than Terzan 5 but still within the uncertainty range.
Tuck \& Lorimer (2013) used an empirical Bayesian approach to determine the possible number of pulsars in GCs.
The results show that Terzan 5, M15, Terzan 6, and NGC 6441 are the four with highest number of pulsars.
The GC 47 Tuc and M28 only rank 10$^{th}$ and 17$^{th}$, respectively.

Among all the studies above,
the predicted number of pulsars in NGC 6517 varies a lot when compared with GCs that have similar known pulsars.
Bagchi et al. (2011) predicted the number of pulsars with different simulation parameters varies between 23$\pm$12 and 271$\pm$136,
being at least twice larger than the predicted numbers among GCs with similar known pulsars, e.g.,
M3 (with 4 known pulsars, Hessels et al. 2007, and 2 unpublished new pulsars from FAST) and M5 (with 5 known pulsars, Anderson et al. 1997, Hessels et al. 2007, and 2 unpublished new pulsars from FAST).
In Tuck \& Lorimer (2013), NGC 6517 did not make the twenty GCs with highest predicted numbers of pulsars.

All the 4 previously known pulsars in NGC 6517 were discovered by GBT at a central frequency of 2~GHz with a bandwidth of 800~MHz (Lynch et al. 2011).
The pulsars NGC 6517 A, C, and D are isolated millisecond pulsars with spin periods shorter than 10~ms.
NGC 6517 B has a spin period of 28~ms and lies in a mildly eccentric orbit with a orbital period of $\rm \sim$60 days, eccentricity of 0.038, and a companion mass of 0.38~\Ms.

The Five-hundred-meter Aperture Spherical radio Telescope (FAST; Nan et al.~2011 and Jiang et al.~2019) is performing a GC survey and has produced at least two GC pulsar publications (Pan et al. 2020, Wang et al. 2020).
NGC6517 is one of the promising targets for FAST GC pulsar survey because it is out of the Arecibo sky and is relatively close to the earth (10.6~kpc).
Drift scan observation of NGC 6517 made in the year of 2017 got no new pulsar detection.
Further observations started in June, 2019.

In this paper, we present the FAST discoveries of three new pulsars in NGC 6517.
In Section 2, we describe the candidate sifting method.
Observation and data reduction are presented in Section 3.
Timing results for known and new isolated pulsars in NGC 6517 are shown in Section 4.
Section 5 and 6 are discussion and conclusions, respectively.

\section{A Simple Pulsar Candidate Sifting Pipeline}

Typical pulsar searches could miss weak millisecond pulsar signals with high dispersion measure (DM, e.g., $\ge$ 100 pc cm$^{-3}$).
The reasons can be explained as follows.

1, the Signal-to-Noise Ratio (SNR) of short period pulsars are affected greatly by DM errors cause by the difference between the pulsar signal's DM and the DM values in the pulsar search DM trial plan.
As a result, a faint short periodic signal with high DM may only be in a small number of DM trials.

2, with a given channel bandwidth, the high DM values also broaden pulse widths and weaken the pulse signals, especially for millisecond pulsars.

3, during the candidate sifting, if a candidate with similar period appears only few times, e.g., 3 times, it may be treated as a faint signal and thus ignored,
no matter this is caused by the DM errors, huge DM steps, or the wider pulse that broadened by the high DM values.

As an example, Einstein@Home reprocessing of the Parkes Multibeam Pulsar Survey data has 24 new pulsars discovered (Knispel et al. 2013).
Among these pulsars, there is a high DM pulsar, PSR~J1748-3009, with a spin period of 9.7~ms and a DM of 420~pc~cm$^{-3}$.
Redetecting this pulsar may not be easy with current pulsar search code, e.g., a typical pulsar search procedure based on PulsaR Exploration and Search TOolkit (PRESTO: Ransom 2001, Ransom et al. 2012, Ransom et al. 2013).
Depending on user's settings, its signal is weak that may only be detected once or twice according to the dedispersion plan obtained with $DDplan.py$ in PRESTO,
and thus gets missed by the sifting code, $ACCEL\underline{ }sift.py$.

One solution for this is to use smaller DM steps during the dedispersion.
With smaller DM step, the faint pulsar signal may show up for more times and thus can be designated as a candidate.
While the small DM step also causes more computing time during dedispersion,
it is necessary to estimate if twice the time cost for possible $\rm \sim$10\% or even lower additional pulsar discoveries is worthy.
Since the new pulsar signal actually exists in the original search results but is missed during the sifting,
a modified sifting code may have a chance to pick up the pulsar signals.
Thus, we designed a graphic-based sifting code to show how the SNR values change with the different DM values.

We divide all the candidates from the pulsar search routine (e.g., $accelsearch$ in PRESTO) for one observation into groups by their periods.
Assuming the SNR of a signal with dispersion will reaches the maximum when the DM value is close to the real one,
the SNR to DM value plot for candidates in each group can be used to see the dispersion features.
Based on PRESTO, we used the so called "Sigma" as the SNR for the following analysis.
The pulsar-like signal should show a SNR peak in a none-zero DM value.
Due to this, we named the code JinglePulsar\footnote{https://www.github.com/jinglepulsar}, as the DM to SNR curve resembles the shape of a bell and we have the first successfully test of code around the Christmas.
In tests, we realised that the periods of candidates in one group may also be sightly different.
In order to see the difference clearly, we used one more plot, showing the period distribution of candidates in each group.
During the candidate grouping, we normally use 4 or 5 digits precision.
The candidates from one observation normally were put into hundreds of groups, resulting in hundreds to thousands of plots to be checked.
With other filtering conditions, such as the lower limit for how many times the candidate appears (we use 3) and the lower limit for highest SNR values of candidates in a group (normally 6 to 8),
the number of groups can be only 1\% of the number of candidates.
The candidate in each group with highest SNR value will be checked again by folding the observation data with the DM value and period of the candidate.
Current candidate ranking codes based on machine learning can be used to check the folding results.
Alternatively, for targeted search, such as GC pulsar search, all the plots can be checked by humans.

\begin{figure*}
\centering
    \includegraphics[width=120mm]{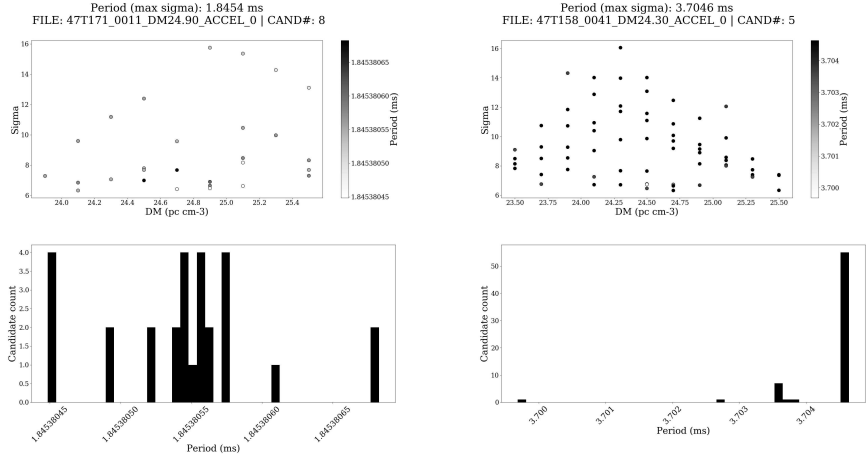}
    \caption{The redetections of 47 Tuc aa (left) and ab (right).
             The upper panels show the relationship between the SNR and DM, the lower panels show the period distributions of the candidate periods.}
    \label{aa_ab}
\end{figure*}

We tested these candidate sifting improvements using the same Parkes observations of 47 Tucanae as described in Pan et al. (2016).
We dedispersed all the data with a DM range of 23.5 to 25.5~pc~cm$^{-3}$ and only searched for isolated pulsars, aiming finding all known isolated pulsars and possible unknown faint pulsars.
There are 313757 raw candidates from the PRESTO search routine $accelsearch$.
These candidates were separated into 2195 groups by their spin period.
After checking the DM to SNR plots by humans, we only selected 48 signals.
These 48 candidates were then confirmed to be 16 known pulsars and related harmonics.
In this test, we successfully detected all the 10 isolated pulsars, including 47 Tuc aa and ab which were discovered by a segmented search in which all the observation data were added incoherently (47 Tuc aa),
or, using PRESTO (47 Tuc ab), respectively.
We also looked up for all the known pulsars those missed during out test in the data.
While none of them were found, this proved that we detected all the known pulsars from the 313757 raw candidates.
Figure \ref{aa_ab} shows the SNR to DM plots and spin period distribution plots of 47 Tuc aa and ab from out test.
We also redetected 6 binaries among all the 15 known binary pulsars.
The reason is that they are bright enough to be detected as isolated pulsars when the acceleration of their orbital movements are close to zero in several minutes during observations.
Unfortunately, no new pulsars were discovered.

\section{Observation and Data Reduction}

In FAST sky, we aim to select GCs with high DM values and known pulsars.
The known pulsars in it can be used for the sifting tests even there is no new pulsar discovered,
and with the known pulsars, the DM range can be largely decreased, thereby saving dedispersion time.
NGC 6517, NGC 6749, and NGC 6760 are the three GCs chosen in FAST sky, have known pulsars, and pulsars in them have relatively high DM values.
Hosting 4 previously known pulsars, NGC 6517 is the first GC selected for a high DM millisecond pulsar search.

We started the NGC 6517 tracking observation on June 25${th}$, 2019,
as one of targets for the FAST GC pulsar survey, which is part of the FAST SP$^2$ survey (Search of Pulsars in Special Population, Pan et al. 2020).
The FAST 19-beam receiver spans a bandwidth of 1.05 to 1.45~GHz with 4096 channels (channel width 0.122~MHz)
The beam size is $\rm \sim$3 arcminutes and the system temperature is $\rm \sim$24~K (Jiang et al. 2020).
We sampled the data from two polarizations with 8-bit precision every 49.152~$\mu$s.
The first observation was done on June 25$^{th}$, 2019, lasting for half an hour.

\begin{figure*}
\centering
    \includegraphics[width=150mm]{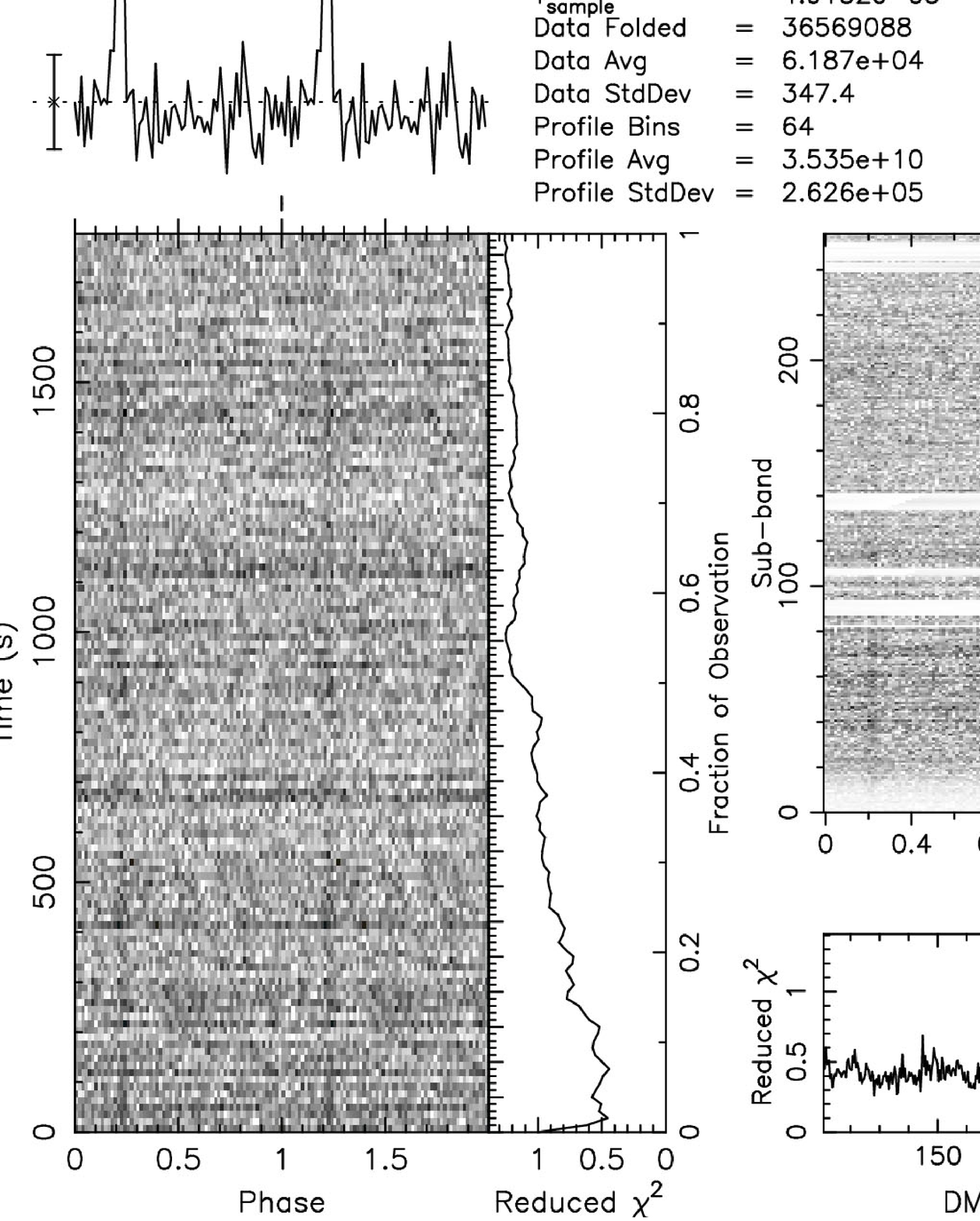}
    \caption{The discovery plots of NGC6517 E, F, and G (from top to bottom), respectively.
             Left panels are the PRESTO folding results of these pulsars and right panels are the DM to SNR plots from our code.
             For the pulsar G, because its DM value is close to the search DM range uplimit, there is not a peak.}
    \label{discovery}
\end{figure*}

The DM from earth to NGC 6517 is $\rm \sim$174 to 183~pc~cm$^{-3}$, determined by its known pulsars.
We dedispersed the data within a DM range of 168.7 to 185.9~pc~cm$^{-3}$.
With a desired time resolution of 0.2~$\mu$s, the DM step calculated by $DDplan.py$ in PRESTO is 0.1~pc~cm$^{-3}$.
In order to search for binary pulsars, the acceleration search was applied by setting the zmax values to be 300.
The search results are sifted by JinglePulsar, results in redetecting all known pulsars and three new signals (showing in Figure \ref{discovery}) which then confirmed to be new pulsars.
The single pulse search was also implemented but no obvious signal was detected.
We also searched the data from the timing observation with same settings, with two more extremely faint isolated pulsars, namely NGC 6517 H and I, discovered.
These two discoveries will be mentioned in another paper (Pan et al. in prep).

FAST can also record baseband data for the 500~MHz bandwidth.
The data are 8-bit sampled for the two polarizations.
Using the 190~pc~cm$^{-3}$ as the DM value, we used coherent dedispersion to create a half-hour observation from the baseband data which we searched as well.
The search detected pulsars A to G.
The SNR from the baseband data is a bit increased, however, yet we detected no new pulsars.

\section{Timing}

So far, a total of 20 observations have been carried out during 447 days between June 2019 (MJD 58659) and September 2020 (MJD 59105).
Depending on the different observation arrangements, observations lasted for less than 0.5 hour to more than 2 hours.
In the data taken on June 25$^{th}$, 2019, we search for best DM values with which the SNR values of pulsars are highest.
The DM values are 182.614, 182.457, 174.487, 183.29, 183.713, and 185.3~pc~cm$^{-3}$ for the pulsar NGC 6517 A, and C to G, respectively,
In each observation, all the pulsars were detected except pulsar G due to its low flux density.
The templates used for TOA generation were made from the longest observation which was taken on January 23$^{rd}$, 2020.
Due to the observation arrangement, the gap between observations can be more than 1 month.
Thus, the method used for determining rotation counts (Freire \& Ridolfi 2018) was used for timing all the six isolated pulsars.
After determining an initial timing solution,
we folded all the observations again with their ephemerides and obtained the improved solutions.
Table \ref{timing_new} shows the timing solutions for the three new pulsars.

As the previously known isolated pulsars NGC 6517 A to D were also detected,
we timed them with the same FAST data.
The timing solution for three isolated pulsars, NGC 6517 A, C, and D, are show on Table \ref{timing_known}.
Comparing with the timing solutions in Lynch et al. (2011), they are consistent.
Due to the high sensitivity of FAST, the timing residuals (4.77 $\mu$s for A, 7.26 $\mu$s for C, and 2.86 $\mu$s for D) are lower than those in Lynch et al. (2011).
For the binary pulsar NGC 6517 B, the gaps in our FAST observations makes us difficult to obtain the timing solution.
We will keep monitoring it and update the timing solution for NGC 6517 B in a future paper.

\begin{table*}
\hspace{-10em}
\caption{Timing solutions for new pulsars NGC6517 E, F, and G.}
\label{timing_new}
\begin{tabular}{cccc}
\hline
                                               & NGC 6517 E            &  NGC 6517 F          &  NGC 6517 G  \\
                                               &  J1801-0857E        &  J1801-0857F       &  J1801-1857G \\
\hline
Number of TOAs                                 & 20                  &  20                &  20         \\
RMS Timing Residual ($\mu$s)                   & 16.87               &  6.96              &  143.3      \\
\hline
\multicolumn{2}{c}{Measured quantities} \\
\hline
Right Ascension, RA (hh:mm:ss, J2000)          &   18:01:50.6232(3)  &  18:01:50.7409(3)  &  18:01:50.099(2)  \\
Declination, DEC (dd:mm:ss, J2000)             &   -08:57:31.29(2)   &  -08:57:31.27(2)   &  -08:57:27.9(1)   \\
Pulse Frequency (Hz)\dotfill                   &  131.55003245964(8) &  40.17357389106(2) &  19.383088017(8) \\
Pulse Frequency Derivative (s$^{-2}$)          & 1.794(2)$\times$10$^{-14}$ & 4.312(7)$\times$10$^{-15}$ & -3(2)$\times$10$^{-17}$  \\
\hline
\multicolumn{2}{c}{Set Quantities} \\
\hline
Reference Epoch (MJD)                         & 58871                &    58871            &  58871        \\
Dispersion Measure (cm$^{-3}$~pc)         & 183.29               &    183.71           &  185.30 \\
\hline
\multicolumn{2}{c}{Timing Model Assumptions} \\
\hline
Solar System Ephemeris Model                  & DE200                &  DE200               &  DE200      \\
\hline
\end{tabular}
\end{table*}

\begin{table*}
\hspace{-20em}
\caption{Timing solutions for perviously known pulsars NGC 6517 A, C, and D.}
\label{timing_known}
\begin{tabular}{cccc}
\hline
                                               & NGC 6517 A            &  NGC 6517 C          &  NGC 6517 D  \\
                                               &  J1801-0857A        &  J1801-0857C       &  J1801-1857D \\
\hline
Number of TOAs                                 & 20                  &  20                &  20         \\
RMS Timing Residual ($\mu$s)                   & 4.77                &  7.26              &  2.86      \\
\hline
\multicolumn{2}{c}{Measured Quantities} \\
\hline
Right Ascension, RA (hh:mm:ss, J2000)          &   18:01:50.6097(1)  &  18:01:50.73731(7)  &  18:01:55.3632(2)  \\
Declination, DEC (dd:mm:ss, J2000)             &   -08:57:31.853(6)  &  -08:57:32.699(3)   &  -08:57:24.284(9)   \\
Pulse Frequency (Hz)\dotfill                   &  139.360884729(3)   &  267.47267494(3)    &  236.600598138(8) \\
Pulse Frequency Derivative (s$^{-2}$)          & 9.9124(8)$\times$10$^{-15}$ & 4.573(9)$\times$10$^{-15}$ & -4.3(2)$\times$10$^{-16}$  \\
\hline
\multicolumn{2}{c}{Set Quantities} \\
\hline
Reference Epoch (MJD)                         & 54400                &    54400            &  54400        \\
Dispersion Measure, DM (cm$^{-3}$~pc)         & 182.614              &    182.457          &  174.487 \\
\hline
\multicolumn{2}{c}{Timing Model Assumptions} \\
\hline
Solar System Ephemeris Model                  & DE200                &  DE200               &  DE200      \\
\hline
\end{tabular}
\end{table*}

\section{Discussion}

The three new pulsars are extremely faint and we believe this is the reason why they have not been discovered by previous studies.
NGC 6517 E and F can be detected by $\rm \sim$0.5-hour FAST observation on L-band, while their fluxes seem to be affected a little by scintillation.
The pulsar G is the faintest among them, and was only detected 16 times among the 20 observations.
Their DM values are all higher than previously known pulsars, indicating that they may be farther away.
Their pulse frequency derivatives are small, showing that they are all recycled pulsars rather than young pulsars, whatever they are millisecond pulsars or relatively long period pulsars.

\begin{figure*}
\centering
    \includegraphics[width=120mm, angle=90]{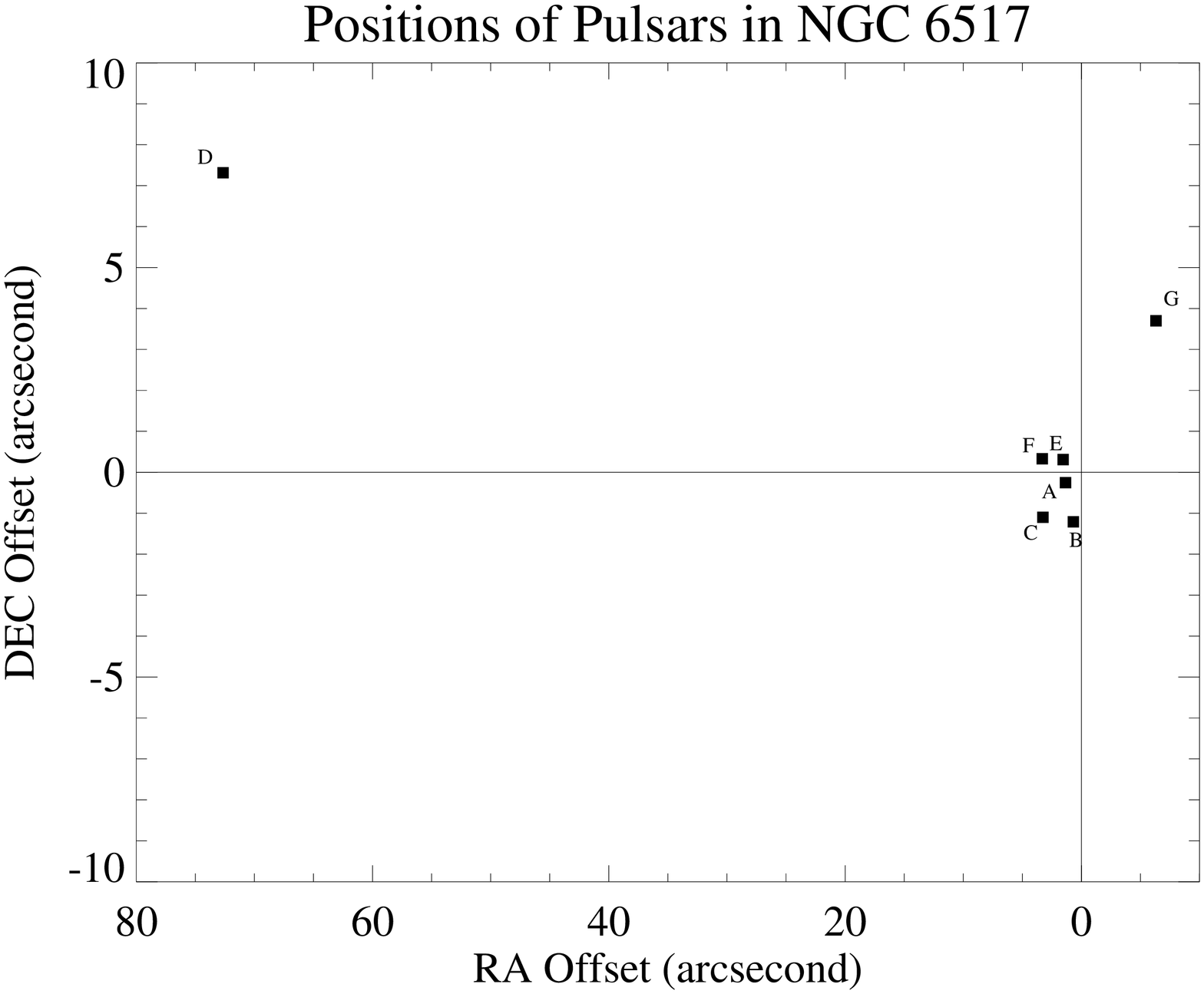}
    \caption{The positions of NGC 6517 A to G.
             The X and Y axes are RA and DEC offsets from the NGC 6517 optical center (J2000, 18:01:50.52 -08:57:31.6, Lynch et al. 2011).
             The error bars for positions are much smaller than the size of squares which represent the pulsars.
             The beam size of FAST is $\rm \sim$3~arcminutes, covers a much larger region than this plot.}
    \label{positions}
\end{figure*}

Figure \ref{positions} shows the positions of all the 7 pulsars in NGC 6517 (B from Lynch et al. 2011).
Pulsar A, B, C, E, and F were very close to each other on the sky and near the GC optical center (J2000, 18:01:50.52 -08:57:31.6, Lynch et al. 2011),
especially for pulsar A, E, and F that there are only a few arcseconds between them.
This case happened rarely in other GC.
The only two possible examples are two pairs of GC pulsars, being 47 Tuc F to S, amd G to I.

With 9 pulsars, NGC 6517 now is the 5$^{th}$ (as same as NGC6752 and M15) of the GCs in terms of numbers of pulsars,
suggesting that previous GC pulsar detection predictions may still have errors.
Our current pulsar surveys also find more pulsar in GCs which are either predicted to have more pulsars or have fewer pulsars.
On the other hand, there is no doubt that the 3 new pulsars were discovered due to better sensitivity.
While we are still looking for new pulsars in other GCs with FAST, Parkes with ultra wideband low frequency receiver, MeerKAT, and other large telescopes,
the relatively high pulsar number of NGC 6517 may be the result of the FAST GC pulsar search priority that we search it earlier.
It is highly possible to find more faint pulsars in other GCs and thus increase the average number of GC pulsars overall.

As seen in Figure \ref{discovery}, for all the three new pulsars, even for the pulsar G which is quite faint,
the SNR to DM curves have clear peaks.
The data points in the curves could be reduce to be half (DM step is 0.2 pc cm$^{-3}$) or even one fourth (DM step is 0.4 pc cm$^{-3}$),
and we could still find these pulsars.
With larger DM steps, the computing time for pulsar search will also be saved.
However, the DM to SNR curve becomes narrower when the pulsar spin period is shorter.
The three new pulsars have relative longer spin periods so that discovered and we concluded the DM step as above.

The DM values for pulsar A, B, C in NGC 6517 are slightly different from Lynch et al. (2011).
It may by either errors from using different band or being caused by the DM variations.
We will continue monitoring this GC and report this in another paper.
While the pulsar NGC 6517 H and I have DM values between NGC 6517 D's and others,
the timing should be done to obtain their positions for further analysis.

With current discoveries, NGC6517 houses 9 pulsars.
Among them, NGC6517B is the only binary and in a $\rm \sim$60 day orbit.
NGC6624 (8 isolated, 1 binary), NGC6752 (8 isolated, 1 binary), M15 (7 isolated, 1 binary, 1 unknown), Terzan 1 (8 islated), and NGC6522 (4 isolated) have similar high isolated pulsar ratios.
NGC6517 also houses 3 long spin period pulsars which have spin period longer than 20 ms.
Except NGC6517 (3 among 9), Terzan 5 (3 among 39), NGC6624 (3 among 10), M71 (3 among 4, discovered by FAST, unpublished), and M15 (4 among 9) have 3 or more pulsars with spin period 20 ms or longer.
Verbunt \& Freire (2014) introduced the encounter rate for a single binary, $\gamma$.
Based on this, they suggested that for core collapse GCs, 
binary pulsar systems have a much higher probability of being destroyed at any stage of the evolution 
and thus these clusters mostly house isolated and/or partially recycled pulsars. 
Thus, the pulsars from NGC6517, M15, and NGC6624 are consistent with the expectation for these core-collapsed GCs

\section{Conclusion}

  The conclusions are as follows.

  1, We discovered 3 new pulsars in NGC 6517.
  With current discoveries, NGC6517 now has 9 pulsars in it, ranking 5$^{th}$ (as same as NGC6752 and M15) of the GCs in terms of numbers of pulsars.

  2, Based on FAST data, we obtained or updated the timing solutions for the 6 isolated pulsars in NGC 6517.
  All these isolated pulsars have small period derivatives, indicating that they are recycled pulsars.

  3, The new candidate sifting code, JinglePulsar,
  was tested on 47 Tuc data and then used for NGC 6517 pulsar search, resulting in the discoveries.
  With these tests, we suggest that with an optimized sifting method,
  the DM step in pulsar search for relatively long period (e.g., 10 ms or longer) pulsars can be at least twice as we used now, with which the computing time can be saved largely.

  4, from our timing solution, the pulsar A, B, C, E, and F are close to each other and near the GC center.

  5, The current GC pulsar searches is performing by large telescope (arrays) like FAST, Parkes UWB, and MeerKAT.
  It will be worthy to re-investigate the simulations on GC pulsar number prediction.

\acknowledgments
This work is supported by the Basic Science Center Project of the National Nature Science Foundation of China (NSFC) under Grant No. 11703047, No. 11773041.
This work made use of data from the Five-hundred-meter Aperture Spherical radio Telescope (FAST).
FAST is a Chinese national mega-science facility, built and operated by the National Astronomical Observatories, Chinese Academy of Sciences (NAOC).
We appreciate all the people from FAST group for their support and assistance during the observations.
The National Radio Astronomy Observatory is a facility of the National Science Foundation operated under cooperative agreement by Associated Universities, Inc.
ZP is supoorted by the CAS "Light of West China" Program.
LQ is supported by the Youth Innovation Promotion Association of CAS (id. 2018075).
SMR is a CIFAR Fellow and is supported by the NSF Physics Frontiers Center award 1430284.

\end{document}